\documentstyle[floats,prl,aps,epsf]{revtex}
\draft % command makes pacs numbers print

\begin{document}

%%%%%%%%%%%%%%%%%%%%%%%%%%%%%%%%%%%%%%%%%%%%%%%%%%%%%%%%%%%%%%%%%%%%%%
% uncomment the following two lines and one below for two columns!
%%%%%%%%%%%%%%%%%%%%%%%%%%%%%%%%%%%%%%%%%%%%%%%%%%%%%%%%%%%%%%%%%%%%%%
\twocolumn[\hsize\textwidth\columnwidth\hsize\csname
@twocolumnfalse\endcsname

\title{The Stability of Relativistic Neutron Stars in Binary Orbit}

\author{T.~W.~Baumgarte$^1$, G.~B.~Cook$^2$, M.~A.~Scheel$^2$, 
	S.~L.~Shapiro$^{1,3}$ and S.~A.~Teukolsky$^{2,4}$}

\address{$^1$Department of Physics, University of Illinois at
        Urbana-Champaign, Urbana, Il~61801}
\address{$^2$Center for Radiophysics and Space Research, Cornell University,
        Ithaca, NY 14853}
\address{$^3$Department of Astronomy and NCSA, University of Illinois at
        Urbana-Champaign, Urbana, Il~61801}
\address{$^4$Departments of Physics and Astronomy, Cornell University,
        Ithaca, NY 14853}
\maketitle

\begin{abstract}
We analyze the stability of relativistic, quasi-equilibrium binary
neutron stars in synchronous circular orbit. We explore stability
against radial collapse to black holes prior to merger, and against
orbital plunge. We apply theorems based on turning points along
uniformly rotating sequences of constant angular momentum and rest
mass to locate the onset of secular instabilities. We find that
inspiraling binary neutron stars are stable against radial collapse to
black holes all the way down to the innermost stable circular orbit.
\end{abstract}


\vskip2pc]

The two-body problem is one of the outstanding, unsolved problems in
classical general relativity.  However, neutron star binary systems
are known to exist, even within our own galaxy~\cite{tw89}.  Binary
systems are among the most promising sources for gravitational wave
detectors now under construction, like LIGO, VIRGO and GEO. This has
motivated an intense theoretical effort to predict the gravitational
wave form emitted during the inspiral and coalescence of two neutron
stars.

Fully general relativistic treatments of the problem are complicated
by the nonlinearity of Einstein's equations and the need for very
large computational resources~\cite{on96}.  Recently, Wilson and
Mathews~\cite{wm95} reported preliminary results obtained with a
relativistic numerical evolution code. Their dynamical calculations
suggest that the neutron stars may collapse to black holes before
their orbit becomes unstable and the stars plunge.  Their results are
in disagreement with predictions of Newtonian~\cite{lrs93},
post-Newtonian (PN)~\cite{l96,w97,lrs97} and perturbation
calculations~\cite{bh97}, which show that tidal fields stabilize
neutron stars against radial collapse.

In a recent paper, we presented quasi-equilibrium, polytropic models
of fully relativistic, equal mass neutron star binaries in
synchronous circular orbits~\cite{bcsst97a}.  In Newtonian gravity,
strict equilibrium exists for two stars in circular orbit. In general
relativity, because of the emission of gravitational waves, binaries
cannot be in strict equilibrium. However, outside of the innermost
stable circular orbit (ISCO), the timescale for orbital decay by
radiation is much longer than the orbital period, so that the binary
can be considered to be in ``quasi-equilibrium''. This fact allows us
to neglect both gravitational waves and wave-induced deviations from a
circular orbit to very good approximation. A detailed discussion of
our approximations and numerical method will be presented in a
forthcoming paper~\cite{bcsst97b}.

In~\cite{bcsst97a}, we focused on the construction of
quasi-equilibrium models, but did not discuss their stability.  We
found that the maximum allowed mass slightly {\em increases} as the
separation of the stars decreases. In this Letter we construct
sequences of constant angular momentum and sequences of constant rest
mass, which then allows us to apply rigorous theorems based on
turning-points along the sequences to locate the onset of secular
instabilities.  We find that neutron stars in synchronous orbit are
stable against radial collapse to black holes until they encounter the
orbital instability at the ISCO.

Applying turning-point methods to binaries in corotation allows us to
detect {\em secular} instabilities.  The separation at which
simultaneous extrema exist in the mass-energy and the angular momentum
curves marks the point along an evolutionary sequence at which the
binary becomes unstable in the presence of some dissipative mechanism.
This instability is distinct from {\em dynamical} instability, which
arises independently of any dissipation and grows on a dynamical
timescale. Gravitational radiation can drive the secular
instability~\cite{c70}, which is expected to occur before (i.e.~at
larger separation than) the onset of a dynamical
instability~\cite{lrs93,fis88}.  It is therefore anticipated that it
is the secular instabilities that limit the range of stable
configurations.

Because of finite numerical resources, we cannot construct models of
very large separation. We therefore observe the increase of the
maximum mass only for separations smaller than about 4 stellar
radii. However, even at this separation, tidal effects are still fairly
small.  Moreover, PN calculations, which are valid for larger
separations, suggest that the maximum mass should also increase
monotonically with decreasing separation~\cite{l96,lrs97}.  In the
following we will therefore assume that the maximum allowed mass
monotonically increases as the separation decreases.

For a given equation of state, cold equal mass binary neutron stars in
synchronous circular orbit form a two-parameter family, just like
single, uniformly rotating stars. In our numerical code we adopt the
central density and relative separation to uniquely specify a
particular configuration. A Lemma by Friedman, Ipser and
Sorkin~\cite{fis88}, originally derived for single, uniformly rotating
stars, can therefore be adapted to corotating binaries:

{\sc LEMMA.} {\em Consider a two-parameter family of equal mass binary
stars in synchronous circular orbits based on an equation of state of
the form $P = P(\rho_0)$. Suppose that along a continuous sequence of
models labeled by a parameter $\lambda$, there is a point $\lambda_0$
at which both $\dot M_0 \equiv dM_0/d\lambda$ and $\dot J$ vanish and
where $d(\dot \Omega \dot J + \dot \mu \dot M_0)/d\lambda \neq 0$.
Then the part of the sequence for which $(\dot \Omega \dot J + \dot
\mu \dot M_0) > 0$ is unstable for $\lambda$ near $\lambda_0$.}

Here $P$ is the pressure, $\rho_0$ the rest-mass density, $J$ and
$M_0$ are the angular momentum and rest (baryon) mass of one of the
two stars, $\Omega$ the angular velocity and $\mu$ is the chemical
potential.  We will later use $M$ for the total mass-energy of one of
the two stars (that is half the total ADM mass of the spacetime).  The
proof of the Lemma follows directly from Theorem 1 of~\cite{s82}. In
practice it is more useful to apply the following related Theorem:

{\sc THEOREM 1.} {\em Consider a continuous sequence of equal mass
binary stars in synchronous circular orbits based on an equation of
state of the form $P = P(\rho_0)$.  Suppose that the total angular
momentum is constant along the sequence, and that there is a point
$\lambda_0$ where $\dot M_0 = 0$ (and where $\mu >
0$, $d(\dot \mu \dot M_0)/d\lambda \neq 0$). Then the part of the
sequence for which $\dot \mu \dot M_0 >0$ is unstable for $\lambda$ near
$\lambda_0$.}

The proof follows directly from the Lemma (see also~\cite{fis88}).

We can now apply this Theorem to our numerical models. In these
models the matter obeys a polytropic equation of state, $ P = K
\rho_0^{1 + 1/n}$,  where $K$ is constant and $n$ is the polytropic
index.  Note that physical dimensions enter the problem only through
$K$. It is therefore convenient to introduce the dimensionless
quantities $\bar \rho_0 = K^n \rho_0$, $\bar M_0 = K^{-n/2} M_0$,
$\bar M = K^{-n/2} M$ and $\bar J = K^{-n} J$.

\begin{figure}
\epsfxsize=3in
\begin{center}
\leavevmode
\epsffile{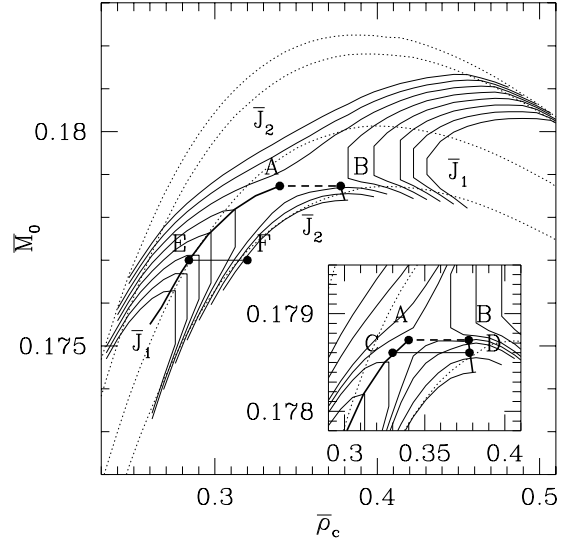}
\end{center}
\caption{Contours of constant angular momentum in a rest-mass versus
central density diagram. The dotted lines are sequences of constant
relative separation $z_A$, ranging from 0.3 (bottom curve) to 0 
(touching stars, top curve). The thin solid lines are contours of
constant angular momentum, ranging from $\bar J_1 = 0.049$ to
$\bar J_2 = 0.0498$ in increments of 0.0001. In the insert we also
show contours for $\bar J = 0.04955$ and 0.049575. The thick solid
lines mark the onset of secular instabilities (see text).}
\end{figure}

In Figure~1 we plot $\bar M_0$ as a function of the central density
$\bar \rho_{\rm c}$ for $n=1$. The dotted lines are curves of constant
relative separation $z_A = r_{\rm in}/r_{\rm out}$. Here $r_{\rm in}$
is the inner (coordinate) distance between the two stars, as measured
between the two closest points along the axis, and $r_{\rm out}$ is
the outer distance, as measured between the most distant points along
the axis. For stars in contact $z_A = 0$, and for infinitely separated
stars $z_A = 1$. The lowest dotted line in Figure 1 is for stars at a
separation $z_A = 0.3$, and the highest curve is for stars in
contact. We expect that the Oppenheimer-Volkoff curve for infinitely
separated stars lies below all these lines~\cite{fn}, such that the
equilibrium mass supported by a given central density monotonically
increases with decreasing separation.

Evolutionary inspiral sequences conserve the rest mass $\bar M_0$ and
therefore follow horizontal lines in Figure~1. This shows, for
example, that the central density $\bar \rho_0$ decreases as the two
stars approach each other.

The solid lines in Figure~1 are contours of constant total angular
momentum $\bar J$. The angular momentum of infinitely separated
binaries is infinite, and decreases for approaching binaries. There is
a critical angular momentum $\bar J^{\rm crit}$, for which the
contours form a saddle point (point A). This critical angular momentum
and the critical mass $\bar M_0^{\rm crit}$ at the saddle point take
values, which, like the maximum mass of nonrotating isolated stars,
depend only on the polytropic index. For $n=1$ we find $\bar J^{\rm
crit} \simeq 0.05$ and $\bar M_0^{\rm crit} \simeq 0.179$~\cite{fn}.

The saddle point divides the turning points of the contours into two
separate classes: one for contours $\bar J < \bar J^{\rm crit}$, and
one for $\bar J > \bar J^{\rm crit}$. We will identify the former with
the instability at the ISCO, and the later with a radial instability
of supramassive stars at large separations.

The class for $\bar J < \bar J^{\rm crit}$ exists for all rest masses
$\bar M_0 < \bar M_0^{\rm crit}$ and ends at the saddle point A. The
physical interpretation of these turning points can be understood by
following an evolutionary sequence for a coalescing binary of fixed
rest mass, which starts at a large separation, e.g., at point F. It
evolves along a horizontal line, and initially both the angular
momentum and the central density decrease with the separation.  At
point E, however, the angular momentum goes through a minimum.
According to the relation~\cite{hb}
\begin{equation} \label{firstlaw}
dM = \Omega dJ + \mu dM_0,
\end{equation}
this minimum has to coincide with an extremum in the
total mass-energy $\bar M$. This extremum is well understood in Newtonian
gravity (e.g.~\cite{lrs93}), PN theory (e.g.~\cite{s96}),
as well as general relativity~\cite{bcsst97a}, and marks the onset of
an orbital instability at the ISCO. Note that this orbital instability is
the first instability that these binaries encounter during inspiral.

For $\bar J > \bar J^{\rm crit}$ we find a second class of turning
points.  This class starts at point B in Figure~1 and ends at the
maximum of the OV curve for isolated, nonrotating stars~\cite{fn,fn2}.
The maximum on this limiting curve marks the onset of radial
instability for isolated, nonrotating stars, and defines their
maximum allowed rest mass $\bar M_0^{\rm max}$. By continuity, this
class thus marks the onset of radial instability for supramassive
sequences, consisting of stars with rest-masses $\bar M_0^{\rm max} <
\bar M_0 < \bar M_0^{\rm crit}$.  Consider, for example, a binary
somewhere on the line between points C and D. 
 Following the inspiral of these stars, they will
eventually incounter the ISCO at point C. If we could reverse the
inspiral and follow an evolutionary sequence towards larger
separations, we know that the stars eventually have to become unstable
and collapse to black holes, since their masses cannot be supported
when in isolation. This will happen at the turning point D. 

Note that so far we have only located turning points of contours of
constant angular momentum. In order to meet all the conditions of
Theorem 1 and rigorously establish the onset of a secular instability,
we also need to examine the chemical potential $\mu$.  The chemical
potential is not normally evaluated by a code that constructs
equilibrium models (although it could be). Accordingly, it is useful
to apply a different version of the Theorem that does not refer to
$\mu$:

{\sc THEOREM 2.} {\em Consider a continuous sequence of equal mass
binary stars in synchronous circular orbits based on an equation of
state of the form $P = P(\rho_0)$.  Suppose that the rest mass $M_0$
is constant along the sequence, and that there is a point
$\lambda_0$ where $\dot M = 0$ (and where $\Omega >
0$, $d(\dot \Omega \dot M)/d\lambda \neq 0$). Then the part of the
sequence for which $\dot \Omega \dot M >0$ is unstable for $\lambda$ near
$\lambda_0$.}

The proof follows from the Lemma together with
relation~(\ref{firstlaw}) (see also~\cite{cst92}). Note that according
to~(\ref{firstlaw}) the extrema in Theorem 1 and 2 have to coincide.

\begin{figure}
\epsfxsize=3in
\begin{center}
\leavevmode
\epsffile{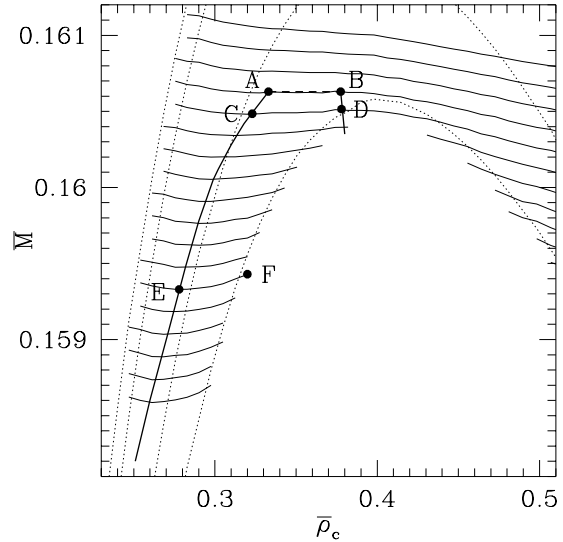}
\end{center}
\caption{Contours of constant rest mass in a mass-energy 
versus central density diagram. The dotted lines are sequences of constant
relative separation $z_A$, ranging from 0.3 (bottom curve) to 0 
(stars in contact, top curve). The thin solid lines are contours of
constant rest mass, ranging from $\bar M_0 = 0.176$ to
$0.1794$ in increments of 0.0002. The thick solid
lines mark the onset of secular instabilities (see text).}
\end{figure}

In Figure 2 we plot contours of constant rest mass $\bar M_0$ in an
$\bar M$ versus $\bar \rho_{\rm c}$ diagram. In this plot the
contours describe secular evolutionary sequences. The contours again
have two qualitatively different classes of extrema. One class are
minima of the total energy (eg.~point C), and the other are maxima
(point D). The former can again be associated with the onset of an
orbital instability at the ISCO, while the later mark the onset of
radial instability of supramassive sequences at large separations.
Both extrema end at the critical contour $\bar M_0^{\rm crit}$, above
which the contours are monotonically decreasing for increasing
$\rho_{\rm c}$. The value of $\bar M_0^{\rm crit}$ agrees with the
height of the saddle point~A and point~B in Figure~1. All extrema in
Figures~1 and~2 agree to within a few percent, which is a measure of
the accuracy of our numerical code.

In order to verify the assumptions in Theorem 2, we have to make sure
that the extrema of the constant rest mass contours do not coincide
with an extremum of the angular velocity. In Figure 3 we show the two
contours connecting points E and F, and C and D, together with 
corresponding plots of $\bar \Omega$. 

The minima at E and C can by analyzed very easily. Using central
density as the parameter $\lambda$, all the assumptions, $\dot M = 0$,
$\Omega > 0$ and $d(\dot \Omega \dot M)/d\lambda \neq 0$ are
satisfied, and we conclude that the configurations to the left of the
minima, for which $\dot \Omega \dot M >0$, are unstable.  These minima
mark the onset of a secular instability close to the ISCO.

The maximum at D is a little less obvious from our numerical data,
since it occurs very close to a minimum in $\bar \Omega$. However, the
offset is measureable and larger than our numerical error, and we
consistently find this offset. In addition, we know that supramassive
stars must become unstable when moved out from finite separation to
infinite separation. We therefore conclude that the extrema in $\bar M$
and $\bar \Omega$ are indeed distinct, so that we can apply Theorem
2. This establishes the onset of a secular radial instability for the
maxima on the thick line extending downward from point B in Figures 1
and 2.

We conclude that all binary configurations in the area under the thick
solid lines and dashes in Figures 1 and 2 are stable. This area can be
subdivided into a normal and a supramassive region. All sequences with
rest masses below the maximum allowed rest mass $\bar M_0^{\rm max}$
of an isolated, nonrotating star are normal sequences. These sequences
start at finite separation (eg.~F) and end at the ISCO (eg.~E).  In
addition to these normal sequences, there are supramassive sequences
with rest masses between $\bar M_0^{\rm max}$ and $\bar M_0^{\rm
crit}$. Evolutionary curves for these sequences connect a radial
instability at large separation (eg.~D) with an orbital instability at
the ISCO (eg.~C).  We conclude that {\em all normal binary neutron
stars in synchronous orbit, with rest masses below the maximum for
isolated, nonrotating configurations, are secularly stable against
radial collapse to black holes down to the ISCO.}  At the ISCO they
encounter a secular orbital instability.

Supramassive stars are unstable to collapse at large separation, but,
if formed at finite separation, remain stable as they inspiral down to
the ISCO.  Supramassive stars could arise in principle if a rapidly
rotating, supramassive core were to undergo bifurcation at finite
radius. Alternatively, they could arise if two spinning neutron stars
become bound in a binary, or if two neutron stars accrete matter while
in a binary.

\begin{figure}
\epsfxsize=2.5in
\begin{center}
\leavevmode
\epsffile{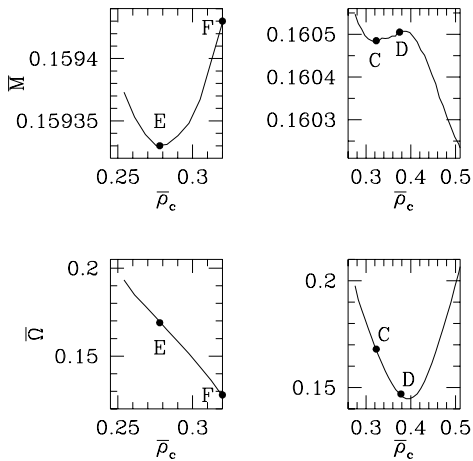}
\end{center}
\caption{Contours of constant rest mass. The top panels show two contours
from Figure~2, and the bottom panels show the corresponding angular
velocities.}
\end{figure}

The secular instability evolves on the same timescale as the inspiral,
since both are driven by gravitational radiation~\cite{lrs93,c70}.
Orbital plunge will occur at the onset of dynamical instability, which
defines the true ISCO. Moreover, realistic binaries are more likely to
be closer to irrotational (circulation $=0$) than
corotational~\cite{viscosity}.  For such systems, however, the
location of the dynamical instability is quite close to the onset of
the secular instability, both in Newtonian theory~\cite{lrs93} and in
PN theory~\cite{lrs97}. The onset of the dynamical instability is at
slightly smaller separation.

The fact that dynamical instabilities typically arise at smaller
separation than secular instabilities, together with the absence of a
secular instability against radial collapse, up to the ISCO, suggests
that a dynamical instability against radial collapse may also be
absent in normal binary neutron stars.

This work was supported by NSF Grant AST 96-18524 and NASA Grant NAG
5-3420 at Illinois, and NSF Grant PHY 94-08378 at Cornell.


\begin{references}
\bibitem{fn2} It is difficult to precisely locate
	the extrema near the saddle point with our numerical
	data. It may very well be that A and B actually coincide.
	However, this would not affect any of our conclusions about stability.
\end{references}
\end{document}